\documentclass[11pt]{article}
\usepackage{mathrsfs}

\usepackage{amsmath}
\usepackage{hyperref}
\usepackage{amsfonts}
\usepackage{amssymb, amsmath, cite}
\usepackage{color}
\usepackage{graphicx}
\usepackage{array}
\usepackage{booktabs}
\usepackage{makecell}
\usepackage{diagbox}
\usepackage[T1]{fontenc}
\usepackage{tabularx}
\usepackage{amsthm}

\numberwithin{equation}{section}
\numberwithin{figure}{section}

\newtheorem{theorem}{Theorem}

\newcommand\be{\begin{equation}}
\newcommand\ee{\end{equation}}
\newcommand\ber{\begin{eqnarray}}
\newcommand\eer{\end{eqnarray}}
\newcommand\berr{\begin{eqnarray*}}
\newcommand\eerr{\end{eqnarray*}}
\newcommand\bea{\begin{eqnarray}}
\newcommand\eea{\end{eqnarray}}

\newtheorem*{theorem*}{Theorem}

\newcommand{\bfR}{{\Bbb R}}\newcommand{\ep}{\epsilon}

\newcommand{\x}{{\bf x}}

\newcommand{\xx}{{\bf x}}

\newcommand{\dd}{\mbox{d}}
\newcommand{\e}{\mbox{e}}
\newcommand{\pa}{\partial}

\newcommand{\nn}{\nonumber}

\newcommand\lb{\label}
\newcommand\eq{\eqref}

\title{The Effective Radius of an Electric Point Charge\\ in Nonlinear Electrodynamics}

\author{{Tengyang Liu}$^{a,}$\footnote{Email address: tyl619@henu.edu.cn} { and Yisong Yang}$^{b,}$\footnote{Email address: yisongyang@nyu.edu}\\[2mm]{\it\small $^a$College of Mathematics and Statistics, Henan University}\\[2mm]{\it\small Kaifeng 475001, P.R.China}\\[2mm]{\it\small $^b$Courant Institute of Mathematical Sciences}\\{\it\small New York University}\\{\it\small New York, New York 10012, USA}}
\date{}

\begin{document}

\maketitle

\begin{abstract}

We develop a new class of nonlinear perturbations of Maxwell’s electrodynamics inspired by, but distinct from, the Born--Infeld theory. A hallmark of our construction is that the effective radius of an electric point charge can be reduced arbitrarily by tuning a coupling parameter, thereby achieving scales far below the Born--Infeld bound and consistent with the experimentally undetected size of the electron. The models preserve finite self energy for point charges while energetically excluding monopoles and dyons, a robustness that appears intrinsic to this class of nonlinear theories. Two complementary behaviors are uncovered: In the 
non-polynomial perturbations, the Maxwell limit is not recovered as the coupling vanishes, whereas in polynomial models the self energy diverges correctly, meaning that  the Maxwellian ultraviolet structure is reinstated. Furthermore, we show that the measurable free electric charge and the self energy contained within any ball around the point charge tend to zero in the strong-nonlinearity or
zero effective radius limit, rendering a pointlike structure locally undetectable, electrically and energetically. These findings highlight how nonlinear field equations reconcile theoretical prescription with experimental measurement and suggest a classical rationale for the effective invisibility of the electron substructure.


\medskip

\medskip




\medskip

\medskip

\medskip

\end{abstract}

\section{Introduction}

The classical problem of modeling the electron as a pointlike particle with finite electromagnetic self energy has long motivated numerous exploration of theories of electrodynamics,
both linear and nonlinear. The most celebrated nonlinear example is the Born--Infeld electrodynamics \cite{B1,B2,BI1,BI2}, which regularizes the Coulomb singularity and yields finite energy for point charges. Yet the Born--Infeld theory comes with a limitation: The effective electron radius it predicts is many orders of magnitude larger than the bounds established by scattering experiments. This discrepancy has presented a challenge and fueled continuing interest in alternative models that might reconcile the point-particle idealization with empirical observations.

In this paper we develop a new class of nonlinear perturbations of Maxwell’s theory that addresses this challenge. Our construction differs from that of Born and Infeld in a crucial respect: While preserving finite self energy, it allows the effective radius of a point charge to shrink arbitrarily by tuning a coupling parameter. This flexibility makes it possible, at least in principle, to push the effective radius below any experimental scale, offering a purely classical rationale for the fact that no substructure of the electron has been detected in the laboratory. At the same time, these models energetically exclude magnetic monopoles and dyons, a structural feature that echoes earlier findings in interpolating the Maxwell and
Born--Infeld theories \cite{LY} and appears robust within this nonlinear framework.

Two complementary families of models emerge from our analysis. The non-polynomial perturbations produce finite self energy and vanishing effective radius, but they do not recover Maxwell’s theory in the weak-coupling limit. In contrast, the polynomial models share the property of arbitrarily small effective radius while restoring the Maxwellian divergence as the coupling parameter tends to zero. This dichotomy reveals a deep asymmetry between charge and energy distributions: While the global energy-charge-radius relation remains valid,
that is, the energy is proportional to the squared charge and inversely proportional to the effective radius, locally the nonlinearities redistribute the free charge and energy in nontrivial ways.

A subtlety that further divides these two types of the models is that the effective radius in the non-polynomial perturbation theories plays the role of a critical scale that effectively borders
distinctive electric and energetic properties of an electric point charge due to the presence of a cutoff threshold and that such a critical behavior disappears in the polynomial perturbation
theories due to the absence of a cutoff threshold. This result associates some clear physical significance to and offers an interpretation of the concept of the effective radius of a point charge introduced by Born and Infeld \cite{BI2} purely as a length scale almost a century ago. 

A further subtlety lies in the distinction between the prescribed source charge $q$, imposed through the displacement field, and the measurable free charge $q_{\rm{free}}$, defined by the divergence of the induced electric field. While the two coincide globally, only a portion of the prescribed charge is contained in a ball around the point charge. Remarkably, in the
 strong-nonlinearity or zero effective radius limit, this local free charge as well as the associated self energy tends to zero, rendering a pointlike structure practically and effectively invisible to measurements. This conclusion provides a classical field-theoretic explanation for the experimentally undetected electron radius, thereby deepening the dialogue between nonlinear electrodynamics and empirical physics.

The rest of the paper is organized as follows. In \S \ref{S2a}, we present a historical review on the point-charge problem, in light of modeling the electron,
that relates to and inspires our work. 
In \S  \ref{S2}, we present the simplest model to be formulated that fulfills our purpose. In \S \ref{S3}, we construct as a general
layout the solutions to the static equations of motion subject to
continuously distributed electric and magnetic charges. In \S \ref{S4}, we consider the electric field generated from a point charge and its properties. Especially we derive the
critical properties of the effective radius of a point charge in Maxwell's limit, both electrically and energetically.
In \S \ref{S5}, we compare in our context the properties of the solutions
obtained here with those of various previously developed models. In particular, we show that those models cannot achieve small effective radius results for an electric point charge.
We also show that our model belongs to a larger family of models of the same properties. A notable feature of this line of construction is that the energy of a point charge
remains finite even in the Maxwell theory limit. That is, we {\em prove} that the theory will {\em not} return to the linear Maxwell theory in the zero nonlinearity limit.  In contrast, in \S \ref{S6}, we show that the polynomial
model does return to the Maxwell theory limit in the description of an electric point charge in the sense that the free electric charge in
any ball around the point charge tends to the prescribed point charge and that the associated self energy in the ball blows up in the zero coupling limit. Meanwhile, the polynomial models also yield an arbitrarily small effective radius for a point charge such that the free electric charge and the self energy contained in any ball around a point charge
tend to zero in the strong coupling or small effective radius limit as in the non-polynomial perturbation theory situations. In \S \ref{S7}, we summarize our results with conclusions and comments.

\section{Historical review and comments}\lb{S2a}

To motivate our study and also to put our study in perspectives, here, we present a short review on our main problem of concern: modeling the electron as a point charge.

In classical electrodynamics or the Maxwell theory, a uniformly charged solid sphere of radius $R$ and total charge $Q$ is calculated to carry an amount of electrostatic self energy, $E$,
which is the energy needed to assemble the charge distribution from infinity and given by the famous formula\cite{Maxwell1,LL,Jackson,Grif}
\be\lb{01}
E=\frac{3}{5}\,\frac{Q^2}{R},
\ee
 in Gaussian units. 
Before the quantum theory of the electron was established, physicists tried to model the electron as a uniformly charged solid ball of charge $e$, the electron charge,  and estimate its size or radius $r_e$ by equating its electrostatic self energy to its rest mass energy $m_e$, the electron mass. Thus, in view of \eq{01}, one arrives at \cite{LL,Jackson,Grif}
\be\lb{02}
r_e =\frac{3}{5} \frac{e^2}{m_e}\approx \frac35\times 2.82 \times 10^{-13}\rm {cm},
\ee
 which is the so-called classical electron radius.
Historically and chronologically, the above idea of estimating the electron radius in terms of the electron mass ---  identified with its electrostatic self energy, and the electron charge, was developed in the late 19th and early 20th centuries by people such as Lorentz\cite{Lorentz1}, Abraham\cite{Abraham1}, and  Poincar\'{e}\cite{Poincar1}. However, modern high-energy scattering experiments and precision tests find no internal structure of the electron down to extremely small distances of the order $10^{-20}$ cm \cite{DDG,HK}, far smaller than $r_e$ given in \eq{02}, leading to an inconsistency,
because the classical picture of a finite-size charged ball renders a much bigger estimate of the electron radius which is not supported experimentally. Furthermore, finding no internal structure of the electron down to all probable scales so far could  mean that the electron behaves as a pointlike particle practically \cite{Dirac,Ro,Te,Yag}.  Unfortunately, a pointlike model would yield a divergent electrostatic self energy\cite{Lorentz1,Maxwell1}. This problematic issue motivated Born and Infeld to propose a nonlinear theory of electrodynamics described by the Lagrangian action density\cite{B1,B2,BI1,BI2}
\be\lb{03}
{\cal L}=b^2\left(1-\sqrt{1-\frac2{b^2}s}\right),\quad s=-\frac14 F_{\mu\nu}F^{\mu\nu}+\frac1{32 b^2}(F_{\mu\nu}\tilde{F}^{\mu\nu})^2,
\ee
where $F_{\mu\nu}$ is the electromagnetic field tensor, $\tilde{F}^{\mu\nu}$ its dual, and $b>0$ a new fundamental field strength scale called the Born parameter.
In the Born--Infeld theory, the source of the electron is still taken to be a point charge concentrated at the space origin and defined by the diverging electric displacement field
but the nonlinear field equations drastically modify the electric field near the origin, such that
 instead of diverging, it saturates at its maximum value $b$.
  This mechanism cures the point charge divergence problem, so that the self energy of a point charge becomes finite.
As a result, even though the source is still a point, the nonlinear field equations make the electric field be distributed continuously over the full space
such that  its main portion of  electric charge is contained in a finite ball of an effective radius,
 $a$. Specifically, following the calculation of Born and Infeld, the total charge of the electric field distributed over the {\em full} space, called the free charge, is the same as the electric charge
$e$.
In the electron model case, the charge $e$ is assigned to the electric displacement field in a pointwise manner and the free charge and the electrostatic self energy or mass contained in the ball of radius $a$ are about 70\% and 60\% of the total charge and mass, respectively\cite{Y-book}.
In this description, the quantity $a$ serves as an important characteristic length scale and is determined to be \cite{BI2}
\be\lb{04}
a=\sqrt{\frac eb},
\ee 
estimated to be about $2.28\times 10^{-13}$ cm and  giving rise to a realization that the Born–Infeld pointlike electron “acts as if” it were an extended object of effective radius $a$.
It is interesting that this estimate is still greater than the classical electron radius stated in \eq{02}. Thus the observed electron radius scale $10^{-20}$ cm has remained unattainable with regard
to the Born--Infeld effective electron radius.

On the other hand, in the last three decades or so, generalized Born--Infeld theories have become an actively pursued area of theoretical physics and have led to fruitful progress in the understanding of
 several important issues including superstring mechanisms \cite{FTs,Lei,Ts}, regularized charged black hole singularities \cite{Ayon1,Ayon2,Kruglov1,Kruglov2,yang2022electromagnetic,YangDyonically,Y-book}, and k-essence cosmology\cite{yang2022electromagnetic,Y-book,CGQ,Linder,Liu}. These theories are collectively governed by the generalized Lagrangian action density of the form \cite{Y-book,Jim}
\be\lb{05}
{\cal L}=f(s),\quad f(0)=0,\quad f'(0)=1,
\ee
where $s$ is given as in \eq{03}.

The main contribution of the present work is to obtain some families of nonlinear theories of electrodynamics of the type \eq{05} which may be used to make the effective radius of a point electric charge
arbitrarily small when an adjustable parameter is being fine-tuned. In particular,  this construction enables us to achieve the length scale below $10^{-20}$ cm for the effective radius of the electron. Besides, we show that, in such a process, the point charge locally loses it characterization as a point charge, which could explain the invisibility of the electron in
view of the electric field it carries.

We emphasize that an important common and distinctive feature of our formalism is that in all models monopoles and dyons are excluded energetically as in our earlier work \cite{LY}
based on a theory that interpolates the Maxwell and Born--Infeld theories, due to the
presence of the Maxwell term in the Lagrangian action density of the theory.

\section{Field-theoretical formulation of the model}\label{S2}
\setcounter{equation}{0}

The main essence of our field-theoretical formalism is centered around a theory of electrodynamics which is a nonlinearly perturbed Maxwell theory based on the Lagrangian action density
\be\lb{1.1}
{\cal L}=f(s)=s+\frac\gamma\beta\left(1-\sqrt{1-(2\beta s)^3}\right),\quad s=-\frac14 F_{\mu\nu}F^{\mu\nu},
\ee
where $F_{\mu\nu}=\pa_\mu A_\nu-\pa_\nu A_\mu$ is the electromagnetic field generated from a real-valued gauge field $A_\mu$ and
the spacetime is equipped with the Minkowski metric $\eta_{\mu\nu}=\mbox{diag}\{1,-1,-1,-1\}$, such that
\be\lb{1.2}
E^i=-F^{0i},\quad B^i=-\frac12\ep^{ijk}F_{jk},\quad i,j,k=1,2,3,
\ee 
are the associated electric and magnetic fields, respectively, and $\beta,\gamma>0$ are coupling parameters, with $\beta$ being related to the Born parameter, through $\beta=b^{-2}$,
and $\gamma$ being dimensionless. It is clear that the function $f(s)$ defined in \eq{1.1} satisfies the conditions  in \eq{05}.

Besides, the variational structure of \eq{1.1} leads to the equations of motion
\be\lb{1.3}
\pa_\mu P^{\mu\nu}=j^\nu,
\ee
where $j^\mu$ is an external source current and
\be\lb{1.4}
P^{\mu\nu}=f'(s) F^{\mu\nu}
\ee
generates the electric displacement field ${\bf D}=(D^i)$ and the magnetic intensity field ${\bf H}=(H^i)$ with
\be\lb{1.5}
D^i=-P^{0i},\quad H^i=-\frac12\ep^{ijk}P_{jk},
\ee
respectively, in analogy to \eq{1.2}. See \cite{Y-book,BB,Ryder}. In view of \eq{1.1}, \eq{1.2}, and \eq{1.5}, we see that \eq{1.4} gives us the constitutive equations 
\bea
{\bf D}&=&f'(s){\bf E}=\left(1+\frac{3\gamma(2\beta s)^2}{\sqrt{1-(2\beta s)^3}}\right){\bf E},\lb{1.6}\\
{\bf H}&=&f'(s){\bf B}=\left(1+\frac{3\gamma (2\beta s)^2}{\sqrt{1-(2\beta s)^3}}\right){\bf B},\lb{1.7}
\eea
with
\be
s=\frac12({\bf E}^2-{\bf B}^2).
\ee
Furthermore, since the energy-momentum tensor formulated from varying the metric tensor $\eta_{\mu\nu}$ in the action integral reads
\be
T_{\mu\nu}=-f'(s)F_{\mu\mu'}\eta^{\mu'\nu'}F_{\nu\nu'}-\eta_{\mu\nu}f(s),
\ee
the Hamiltonian energy density of \eq{1.1} is
\bea\lb{1.9}
{\cal H}&=&T_{00}=f'(s){\bf E}^2-f(s)\nn\\
&=&\left(1+\frac{3\gamma(2\beta s)^2}{\sqrt{1-(2\beta s)^3}}\right){\bf E}^2-\left(s+\frac\gamma\beta\left(1-\sqrt{1-(2\beta s)^3}\right)\right)\nn\\
&=&\frac12({\bf E}^2+{\bf B}^2)+\frac{3\gamma(\beta[{\bf E}^2-{\bf B}^2])^2{\bf E}^2}{\sqrt{1-(\beta [{\bf E}^2-{\bf B}^2])^3}}-\frac{\gamma(\beta [{\bf E}^2-{\bf B}^2])^3}{\beta(1+\sqrt{1-(\beta [{\bf E}^2-{\bf B}^2])^3})},
\eea
which appears complicated. Fortunately, in the electrostatic and magnetostatic situations with ${\bf B}={\bf 0}, {\bf H}={\bf 0}$ and ${\bf D}={\bf 0},
{\bf E}={\bf0}$, respectively, the expression \eq{1.9} simplifies itself greatly and transparently. In fact, in the former situation, we have
\be\lb{a2.10}
{\cal H}=\frac12{\bf E}^2+\frac{\gamma(\beta{\bf E}^2)^3(3+2\sqrt{1-(\beta {\bf E}^2)^3})}{\beta\sqrt{1-(\beta{\bf E}^2)^3}(1+\sqrt{1-(\beta{\bf E}^2)^3})},
\ee
and in the latter situation, we have
\be\lb{a2.11}
{\cal H}=\frac12{\bf B}^2+\frac{\gamma(\beta {\bf B}^2)^3}{\beta(1+\sqrt{1+(\beta{\bf B}^2)^3})},
\ee
both being positive definite. These formulas will be convenient for our subsequent calculations.

The dyonic situation in which both electric and magnetic fields are present will be addressed in the next section.

\section{Static systems subject to arbitrary sources}\label{S3}
\setcounter{equation}{0}

First consider the electrostatic situation.  The constitutive equations, \eq{1.6} and \eq{1.7},  become a single one,
\be\lb{2.1}
{\bf D}=\left(1+\frac{3\gamma(\beta {\bf E}^2)^2}{\sqrt{1-(\beta {\bf E}^2)^3}}\right){\bf E},
\ee
which leads to
\be\lb{2.2}
\beta{\bf D}^2=\left(1+\frac{3\gamma\eta^2}{\sqrt{1-\eta^3}}\right)^2 \eta\equiv g(\eta),\quad \eta=\beta{\bf E}^2.
\ee
It may be checked that $g'(\eta)>0$ for $\eta\in[0,1)$. Hence \eq{2.2} can be inverted to render us
\be\lb{2.3}
\eta=\beta {\bf E}^2\equiv h(\beta{\bf D}^2),
\ee
with the properties
\be\lb{2.4}
 \lim_{{\bf D}^2\to 0}h(\beta{\bf D}^2)=0,\quad  \lim_{{\bf D}^2\to \infty}h(\beta{\bf D}^2)=1,\quad
\lim_{{\bf D}^2\to\infty}\beta {\bf D}^2 (1-h^3(\beta{\bf D}^2))=(3\gamma)^2,
\ee
such that \eq{2.1} may be resolved to yield 
\be\lb{2.5}
{\bf E}=\frac{{\bf D}}{1+\frac{3\gamma h^2(\beta{\bf D}^2)}{\sqrt{1-h^3(\beta{\bf D}^2)}}}.
\ee
By \eq{2.4}, we have
\be\lb{2.7}
\lim_{{\bf D}^2\to\infty}{\bf E}^2=\frac1{\beta}.
\ee
In other words, $\bf E$ stays finite at any blow-up point of $\bf D$ as in the classical Born--Infeld theory \cite{B1,B2,BI1,BI2}.

Let $\rho_e(\x)$ be an electric charge density distribution which generates the electric displacement field $\bf D$ through \eq{1.3}. Then we have
\be\lb{2.12}
\nabla\cdot{\bf D}=\rho_e.
\ee
As a result, the total electric charge is given by
\be\lb{a3.8}
Q_e=\int_{\bfR^3}\rho_e(\x)\,\dd\x=\int_{\bfR^3}\nabla\cdot{\bf D}\,\dd\x=\lim_{r\to\infty}\int_{|\x|=r}{\bf D}\cdot \dd{\bf S}.
\ee
In view of \eq{a3.8}, the finiteness of the total charge leads to the asymptotic property
\be\lb{2.14}
{\bf D}=\mbox{O}(r^{-2}),\quad |\x|=r\gg1,
\ee
or Coulomb's law.
On the other hand, from \eq{2.5}, we have
\bea\lb{2.9}
{\bf E}&=&{\bf D}+\left(\frac1{1+\frac{3\gamma h^2(\beta{\bf D}^2)}{\sqrt{1-h^3(\beta{\bf D}^2)}}}-1\right){{\bf D}}\nn\\
&=&{\bf D}-\frac{3\gamma h^2(\beta{\bf D}^2){\bf D}}{\sqrt{1-h^3(\beta{\bf D}^2)}+3\gamma h^2(\beta{\bf D}^2)}.
\eea
Furthermore, using \eq{2.2}, we get
\be\lb{a3.11}
h'(0^+)=\frac1{g'(0^+)}=1.
\ee
Inserting \eq{a3.11} into \eq{2.9}, we have
\be\lb{2.10}
{\bf E}={\bf D}+\mbox{O}(|{\bf D}|^5),\quad |{\bf D}|\ll1.
\ee
In view of \eq{2.10} and \eq{2.14}, we see that with the definition of the free electric charge density \cite{BI2,Y-book}
\be
\nabla\cdot{\bf E}=\rho_{\rm{free}},
\ee
the total free electric charge is
\bea
Q_{\rm{free}}&=&\int_{\bfR^3}\rho_{\rm{free}}(\x)\,\dd\x=\int_{\bfR^3}\nabla\cdot{\bf E}\,\dd\x=\lim_{r\to\infty}\int_{|\x|=r}{\bf E}\cdot \dd{\bf S}\nn\\
&=&\lim_{r\to\infty}\int_{|\x|=r}{\bf D}\cdot \dd{\bf S}=Q_e.
\eea
That is, the free electric charge distributed in space generated by the electric field $\bf E$ is in agreement with the prescribed total charge $Q_e$ assigned to the electric displacement field $\bf D$ which may either be continuously distributed
or localized at the charge concentration points in space.

Using \eq{2.14}  and \eq{2.10} in \eq{a2.10}, we see that finite charge condition implies finite total energy 
\be\lb{2.17}
E=\int_{\bfR^3}{\cal H}\,\dd\x<\infty,
\ee
in all kinds of electric charge distribution situations, discrete and continuous.

The magnetostatic situation is different, though.

In fact, a similar discussion leads to the conclusion that the free magnetic charge generated from the induced magnetic intensity field $\bf H$ and 
the prescribed magnetic charge given by the magnetic field $\bf B$ are the same,
\be
G_{\rm{free}}=\int\nabla\cdot {\bf H}\,\dd {\bf x}=\int\nabla\cdot{\bf B}\,\dd{\bf x}=\int {\rho}_m\,\dd{\bf x}\equiv G_m,
\ee
where $\rho_m=\rho_m({\bf x})$ is the magnetic charge density distribution which generates the magnetic field $\bf B$ through the magnetic Poisson equation
\be\lb{a3.17}
\nabla\cdot {\bf B}=\rho_m.
\ee
Finiteness of $G_m$ implies that $\bf B$ decays like $\bf D$ before:
\be\lb{a3.18}
{\bf B}=\mbox{O}(r^{-2}),\quad |{\bf x}|=r\gg1.
\ee

Thus, if $\rho_m$ is {\em continuously} distributed such that $\bf B$ remains finite in space, then, inserting \eq{a3.18} into \eq{a2.11}, we arrive at the same
conclusion that the total energy of the magnetostatic field is finite.

\medskip

{\bf Exclusion of monopoles}

On the other hand, however, the discrete magnetic point charge situation does not enjoy such a finiteness property.

To see this, we consider a magnetic point charge $g>0$ placed at the origin so that its charge density $\rho_m$ is given by the Dirac measure:
\be\lb{a3.19}
\rho_m(\x)=4\pi g\delta(\x).
\ee
With \eq{a3.19}, the Poisson equation \eq{a3.17} gives us the magnetic Coulomb law solution:
\be\lb{a3.20}
{\bf B}=\frac{g\x}{r^3},\quad r=|\x|>0.
\ee
In view of \eq{a3.20} and \eq{a2.11}, we see that the total energy of the magnetic field generated from a magnetic point charge is infinite. More precisely, since the energy blow-up occurs at
the origin, we encounter an ultraviolet divergence situation. In other words, energetically, the model \eq{1.1} does not accommodate any monopoles.

\medskip

{\bf Exclusion of dyons}

More generally, we ask whether the model would allow dyons. To this end, we note that \eq{1.9} may be rewritten as
\bea
{\cal H}&=&\frac12({\bf E}^2+{\bf B}^2)+\frac{\gamma(\beta[{\bf E}^2-{\bf B}^2])^2{\bf E}^2(3+2\sqrt{1-(\beta [{\bf E}^2-{\bf B}^2])^3})}{\sqrt{1-(\beta [{\bf E}^2-{\bf B}^2])^3}(1+\sqrt{1-(\beta [{\bf E}^2-{\bf B}^2])^3})}\nn\\
&&+\frac{\gamma(\beta [{\bf E}^2-{\bf B}^2])^2{\bf B}^2}{1+\sqrt{1-(\beta [{\bf E}^2-{\bf B}^2])^3}}.
\eea
Hence, we conclude as in the monopole situation that a dyonic point charge does not carry a finite energy either,  due to \eq{a3.20}, although it is clear that continuously
distributed dyonic charges are of finite energies as before as in the electric and magnetic charge situations \cite{Yang-preprint}.

\medskip

Subsequently, we shall focus on an electric point charge problem.

\section{Electric point charge problem}\label{S4}
\setcounter{equation}{0}

Now we consider a point charge $q>0$ residing at the origin of the space. The electric displacement field $\bf D$ satisfies the Poisson equation 
\eq{2.12} with $\rho_e(\x)=4\pi q\delta(\x)$ such that $\bf D$ is given by Coulomb's law
\be\lb{3.1}
{\bf D}=\frac{q\x}{r^3},\quad r=|\x|,\quad \x\neq{\bf 0}.
\ee
Substituting $\eta=\beta {\bf E}^2$ into \eq{a2.10} and integrating, we have 
\bea\lb{3.6}
\frac E{4\pi}&=&\int_0^\infty {\cal H}\,r^2\,\dd r\nn\\
&=&\frac1{2\beta}\int_0^\infty\left(\eta+\frac{2\gamma\eta^3(3+2\sqrt{1-\eta^3})}{\sqrt{1-\eta^3}(1+\sqrt{1-\eta^3})}\right)r^2\,\dd r\nn\\
&=&\frac1{2\beta}\int_0^\infty\left(h\left(\frac{a^4}{r^4}\right)+\frac{2\gamma h^3\left(\frac{a^4}{r^4}\right)\left(3+2\sqrt{1-h^3\left(\frac{a^4}{r^4}\right)}\right)}{\sqrt{1-h^3\left(\frac{a^4}{r^4}\right)}\left(1+\sqrt{1-h^3\left(\frac{a^4}{r^4}\right)}\right)}\right)r^2\,\dd r\nn\\
&=&\frac{q^2}{2a}\int_0^\infty\left(h\left(\frac{1}{x^4}\right)+\frac{2\gamma h^3\left(\frac{1}{x^4}\right)\left(3+2\sqrt{1-h^3\left(\frac{1}{x^4}\right)}\right)}{\sqrt{1-h^3\left(\frac{1}{x^4}\right)}\left(1+\sqrt{1-h^3\left(\frac{1}{x^4}\right)}\right)}\right)x^2\,\dd x,
\eea
where
\be\lb{axx5.3}
a=\beta^{\frac14}q^{\frac12}
\ee
is identified to be the effective radius of the point charge following the formulation of Born and Infeld \cite{BI2} and we have set $r=ax$ in \eq{3.6}.

To proceed further, insert \eq{3.1} into \eq{2.2} to get
\be\lb{3.8}
x=\left(1+\frac{3\gamma \eta^2}{\sqrt{1-\eta^3}}\right)^{-\frac12}\eta^{-\frac14},\quad \eta=h\left(\frac1{x^4}\right),
\ee
such that $x\to 0$ and $x\to\infty$ correspond to $\eta\to1$ and $\eta\to0$, respectively. 

Using \eq{3.8} in \eq{3.6}, we have
\bea\lb{3.9}
\frac{E}{4\pi}&=&\frac{q^2}{8a}\int_0^1\left(1+\frac{2\gamma\eta^2(3+2\sqrt{1-\eta^3})}{\sqrt{1-\eta^3}(1+\sqrt{1-\eta^3})}\right)\frac{(\gamma\eta^2(15-6\eta^3)+(1-\eta^3)^{\frac32})}{\eta^{\frac34}(1-\eta^3)^{\frac14}(3\gamma\eta^2+\sqrt{1-\eta^3})^{\frac52}}\,\dd \eta\nn\\
&\equiv&\frac{q^2}a\int_0^1 h_\gamma(\eta)\,\dd\eta\nn\\
&\equiv&\frac{q^2}a H(\gamma).
\eea

Although the quantity $H(\gamma)$ does not allow a precise determination through an exact integration, it enjoys the important property
\be\lb{a4.6}
\lim_{\gamma\to\infty}H(\gamma)=0,
\ee
which enables us to achieve the arbitrary smallness for the effective radius of an electric point charge $q$ with a given mass $m$. Indeed, replacing the
left-hand side of \eq{3.9} by $m$ as in \cite{BI2}, we get
\be\lb{a4.7}
a=\frac{q^2}m H(\gamma).
\ee
Hence the asserted arbitrary smallness of the effective radius is established in view of \eq{a4.6}.

Besides, in view of \eq{3.9}, we have 
\be\lb{4.8}
\lim_{\gamma\to0}H(\gamma)=\frac18\int^1_0\frac{\dd\eta}{\eta^{\frac34}}=\frac12,
\ee
which gives us the {\em finite} exact self energy of a {\em point} charge in the Maxwell theory limit
\be\lb{x4.9}
\lim_{\gamma\to0}\frac E{4\pi}=\frac{q^2}{2a},
\ee
which is in interesting comparison with the classical result given in \eq{01}.

As a consequence of \eq{x4.9},  we see that the effective radius of an electric point charge of charge $q$ and mass $m$  in the Maxwell theory limit is given by
\be
a=\frac{q^2}{2m}.
\ee

\medskip

{\bf Free charge distribution of a point charge}

We next study how the free electric charge of an electric point charge distributes itself in space. 

With the electric displacement field ${\bf D}$ given in \eq{3.1}, we see that the radial component of the electric field $\bf E$, say $E_r$, relates
to $D_r=\frac q{r^2}$, by \eq{2.2}, which gives us
\be\lb{a4.8}
\frac{a^4}{r^4}=\left(1+\frac{3\gamma\eta^2}{\sqrt{1-\eta^3}}\right)^2 \eta,\quad \eta=\beta E_r^2.
\ee
On the other hand, recall that in the radial situation, we have
\bea\lb{a4.9}
\rho_{\rm{free}}(r)&=&\nabla\cdot{\bf E}=\frac1{r^2}\frac{\dd}{\dd r}(r^2 E_r)\nn\\
&=&\frac2r\left(\frac\eta\beta\right)^{\frac12}+\frac1{2\beta}\left(\frac\eta\beta\right)^{-\frac12}\frac{\dd\eta}{\dd r}.
\eea
However, differentiating \eq{a4.8}, we have
\be\lb{a4.10}
-\frac{4}r\left(\frac ar\right)^4=\frac{(3\gamma \eta^2+\sqrt{1-\eta^3})(3\gamma\eta^2(5-2\eta^3)+(1-\eta^3)^{\frac32})}{(1-\eta^3)^2}\,\frac{\dd\eta}{\dd r}.
\ee
Inserting \eq{a4.10} into \eq{a4.9}, we obtain
\bea
\rho_{\rm{free}}(r)&=&\frac2{\beta^{\frac12}r}\left(\eta^{\frac12}-\left(\frac ar\right)^4
\frac{(1-\eta^3)^2}{\eta^{\frac12}(3\gamma \eta^2+\sqrt{1-\eta^3})(3\gamma\eta^2(5-2\eta^3)+(1-\eta^3)^{\frac32})}\right)\nn\\
&=&\frac{2q}{a^3\left(\frac ra\right)}\left(\eta^{\frac12}-\left(\frac ar\right)^4
\frac{(1-\eta^3)^2}{\eta^{\frac12}(3\gamma \eta^2+\sqrt{1-\eta^3})(3\gamma\eta^2(5-2\eta^3)+(1-\eta^3)^{\frac32})}\right),
\eea
explicitly given in terms of the effective radius. 

Furthermore, integrating \eq{a4.9} over the ball $\{|\x|\leq r\}$, we have
\bea\lb{xx5.15}
Q_{\rm{free}}(r)&=&4\pi r^2 E_r=4\pi r^2 \left(\frac\eta\beta\right)^{\frac12}\nn\\
&=&4\pi q\left(\frac ra\right)^2\eta^{\frac12},\quad \eta=h\left(\frac {a^4}{r^4}\right).
\eea
Therefore, the free electric charge contained in the ball of  radius $R>0$ around the electric point charge enjoys the simple formula
\be\lb{x4.16}
Q_{\rm{free}}(R)=4\pi q\left(\frac Ra \right)^2 \sqrt{h\left(\frac {a^4}{R^4}\right)}.
\ee
Since the prescribed electric charge is $Q=4\pi q$, we obtain the ratio of the generated free charge over the ball $\{|\x|\leq R\}$ against the prescribed total charge:
\be
\frac{Q_{\rm{free}}(R)}Q=\left(\frac Ra \right)^2 \sqrt{h\left(\frac {a^4}{R^4}\right)}=\left(\frac Ra \right)^2 \eta^{\frac12},
\ee
where $\eta=\eta(\gamma)$ is determined implicitly by \eq{a4.8} after setting $r=R$. That is, $\eta(\gamma)$ is the unique solution to the equation
\be\lb{a4.15}
\left(1+\frac{3\gamma\eta^2}{\sqrt{1-\eta^3}}\right)^2 \eta=\frac{a^4}{R^4}.
\ee
We are unable to solve for $\eta(\gamma)$ in terms of $\gamma$ explicitly in \eq{a4.15} but it is useful to represent $\gamma$ in terms of $\eta$:
\be\lb{a4.16}
\gamma=\frac{\sqrt{1-\eta^3}}{3\eta^{\frac52}}\left(\frac{a^2}{R^2}-\eta^{\frac12}\right).
\ee
It is clear that consistency of this equation leads to
\be\lb{ay5.20}
\eta(\gamma)\leq\min\left\{1,\frac{a^4}{R^4}\right\},
\ee
such that $\eta(\gamma)$ decreases with respect to $\gamma$ and
\be\lb{ay5.21}
\lim_{\gamma\to0}\eta(\gamma)=\frac{a^4}{R^4}\quad (R\geq a),\quad \lim_{\gamma\to0}\eta(\gamma)=1\quad (R\leq a);\quad \lim_{\gamma\to\infty}\eta(\gamma)=0.
\ee

This picture leads us to the following scenarios about the free electric charge generated from the electric field $\bf E$ of an electric point charge given by \eq{3.1}:

\begin{enumerate}

\item[(i)] In the Maxwell theory limit, $\gamma\to0$, the free electric charge contained in the ball of  radius $R$ follows the distributional properties
\bea
&& \lim_{\gamma\to0} Q_{\rm{free}}(R)=4\pi q\left(\frac Ra\right)^2,\quad R\leq a;\lb{axx5.22a}\\
&&\lim_{\gamma\to0} Q_{\rm{free}}(R)=Q=4\pi q,\quad R\geq a. \lb{axx5.22b}
\eea

\item[(ii)] In the strong nonlinearity situation, $\gamma\gg1$, the free electric charge contained in the ball of any radius $R$ takes only
a small portion of the prescribed electric charge and it approaches zero as $\gamma\to\infty$.

\item[(iii)] The coupling parameter $\gamma$ can be chosen to render the free electric charge contained in the ball of the radius $R\geq a$ around the point charge
to be any portion of the prescribed electric charge.

\end{enumerate}

\medskip

{\bf Self energy distribution of a point charge}

Equally interestingly, since the energy contained in the ball of  any radius $R>0$ around the point charge is
\bea\lb{4.20}
E(R)&\equiv&\int_{|\xx|\leq R}{\cal H}\,\dd\xx\nn\\
&=&4\pi\int_0^R {\cal H} r^2\,\dd r\nn\\
&=&\frac{4\pi q^2}a\int^1_{\eta(\gamma)} h_\gamma(\eta)\,\dd\eta,
\eea
then \eq{ay5.21} enables us to draw the conclusions
\bea
&&\lim_{\gamma\to 0} E(R)=0,\quad R\leq a;\lb{axx5.25a} \\
&& \lim_{\gamma\to 0} E(R)=\frac{4\pi q^2}a\int^1_{\frac{a^4}{R^4}}\frac{\dd\eta}{8\eta^{\frac34}}=2\pi q^2\left(\frac1a-\frac1R\right),\quad R\geq a,\lb{axx5.25b}
\eea
regarding the self energy of the electric field, contained in the ball of  radius $R>0$,  generated around an electric point charge, in the Maxwell theory limit.  These results
refine the global result  
 \eq{x4.9} over the full space.

\medskip

The picture depicted by the results \eq{axx5.22a}, \eq{axx5.22b}, \eq{axx5.25a}, and \eq{axx5.25b} regarding the free electric charge and self energy distributions implicates that the energy-charge relation
\be\lb{axx5.24}
\mbox{Energy}\sim\frac{\mbox{Charge}^2}{\mbox{Effective radius}}
\ee
is a global one in the full parameter regime, as stated in \eq{3.9}, and that such a relation is invalid locally  around an electric point charge in the Maxwell
theory limit. That is, we
encounter
 a mass and charge disparity or asymmetry phenomenon in the Born--Infeld type model \eq{1.1}, which reveals the distinct subtle ways the charge and mass respond to field nonlinearities, rather unambiguously.

\section{General formalism and some examples for comparison}\label{S5}
\setcounter{equation}{0}

In a more refined form, the generalized Born--Infeld theory \eq{05}  may be rewritten as
\be\lb{5.2}
{\cal L}=\frac1\beta\, U(\beta s),\quad U(0)=0,\quad U'(0)=1,\quad \beta>0.
\ee

In terms of the function $U$, the constitutive equations \eq{1.6} and \eq{1.7} in the electrostatic situation reduce into the single equation
\be\lb{5.3}
{\bf D}=U'(\beta s){\bf E}.
\ee
Squaring \eq{5.3} and multiplying by $\beta$, we obtain
\be\lb{5.4}
\left(U'\left(\frac{\beta {\bf E}^2}2\right)\right)^2(\beta {\bf E}^2)=\beta{\bf D}^2,
\ee
which may be inverted to give us the formal solution
\be
\eta=\beta{\bf E}^2=h(\beta {\bf D}^2)
\ee
as before. Note that, to ensure the invertibility of \eq{5.4} or
\be\lb{5.6}
\left(U'\left(\frac\eta2\right)\right)^2\eta=\beta{\bf D}^2,
\ee
it suffices to impose the condition
\be\lb{5.7}
2tU'(t)U''(t)+(U'(t))^2>0,\quad t>0,
\ee
in the domain of interest of the function $U(t)$, which will be assumed subsequently.
 Moreover, the Hamiltonian energy density \eq{1.9} reads
\bea\lb{5.8}
{\cal H}&=&U'(\beta s){\bf E}^2-\frac1{\beta}\,U(\beta s)\nn\\
&=&\frac1\beta(2U'(\beta s)(\beta s)-U(\beta s)),\quad s=\frac{{\bf E}^2}2.
\eea
Thus, the positivity condition ${\cal H}\geq0$ implies that we need to impose the global condition
\be\lb{5.9}
2U'(t)t-U(t)\geq0,\quad t\geq0,
\ee
in its domain of interest in addition to the local condition given in \eq{5.2}. 

Under these sufficient conditions and inserting the electric point charge information
\eq{3.1} into \eq{5.6}, we have
\be\lb{5.10}
\left(U'\left(\frac\eta2\right)\right)^2\eta=\frac{a^4}{r^4}=\frac1{x^4}.
\ee
In view of \eq{5.10}, we see that \eq{5.8} leads to
\bea\lb{5.11}
\frac{E}{4\pi}&=&\frac{a^3}{\beta}\int_0^\infty \left(U'\left(\frac\eta2\right)\eta-U\left(\frac\eta2\right)\right)\,x^2\,\dd x\nn\\
&=&\frac{q^2}{a}\int \left(U'\left(\frac\eta2\right)\eta-U\left(\frac\eta2\right)\right)\,\frac{\left( U'\left(\frac\eta2\right)U''\left(\frac\eta2\right)\eta+\left(U'\left(\frac\eta2\right)\right)^2\right)}{4\left(\left(U'\left(\frac\eta2\right)\right)^2\eta\right)^{\frac74}}\,\dd \eta,
\eea
where the interval of integration in the variable $\eta$ is arranged in the increasing direction of $\eta$.

\medskip

{\bf Some examples}

We now consider some known examples in light of this refined formalism where the function $U$ given in \eq{5.2} also satisfies the conditions \eq{5.7} and \eq{5.9} in addition.

(i) The classical Born--Infeld model \cite{B1,B2,BI1,BI2}  is given by
\bea
f(s)&=&\frac1\beta\left(1-\sqrt{1-2\beta s}\right),\label{5.12}\\
U(t)&=&1-\sqrt{1-2t}.\label{5.13}
\eea
From \eq{5.10}, we see that the interval of the variable $\eta$ is $[0,1)$. So, inserting \eq{5.13} into \eq{5.11}, we obtain the classical result
\begin{align}
\frac E{4\pi}&=\frac{q^2}{4a}\int^1_0\left((1-\eta)^{-\frac12}-(1+\sqrt{1-\eta})^{-1}\right)\eta^{-\frac34}(1-\eta)^{-\frac74}{\rm{d}}\eta\nn\\
&\approx1.23605\left(\frac{q^2}{a}\right).
\end{align}

(ii) The binomial model\cite{Costa2015,Lorenci2002,Garcia-Salcedo2014} is defined by
\begin{align}
f(s)=&s+\beta s^2,\label{5.15}\\
U(t)=&t+t^2.\lb{5.15b}
\end{align}
From \eq{5.10}, we see that the interval of the variable $\eta$ is $[0,\infty)$. Thus, inserting \eq{5.15b} into \eq{5.11}, we have
\begin{align}
\frac E{4\pi}&=\frac{q^2}{a}\int^\infty_0\frac{(2+3\eta)(1+3\eta)}{16(1+\eta)^\frac52\eta^\frac34}{\rm{d}}\eta\nn\\
&\approx2.4721\left(\frac{q^2}{a}\right).\label{5.17}
\end{align}

(iii) The exponential model\cite{Hendi2012,Hendi2013} reads
\begin{align}
f(s)=&\,\frac1\beta(\e^{\beta s}-1),\\
U(t)=&\,\e^t-1.\lb{5.18}
\end{align}
From \eq{5.10}, we see that the interval of the variable $\eta$ is $[0,\infty)$. Hence, inserting \eq{5.18} into \eq{5.11}, we obtain
\begin{align}
\frac E{4\pi}&=\frac{q^2}{4a}\int^\infty_0\left(\e^{\frac\eta2}\eta-\e^{\frac\eta2}+1\right)\eta^{-\frac74}\e^{-\frac{3\eta}{4}}(1+\eta){\rm{d}}\eta\nn\\
&\approx1.70913\left(\frac{q^2}{a}\right).
\end{align}

(iv) Another exponential model \cite{Kruglov2017nonlinear} is defined by
\begin{align}
f(s)&= s\e^{\beta s},\\
U(t)&= t\e^t.\lb{5.21}
\end{align}
From \eq{5.10}, we see that the interval of the variable $\eta$ is $[0,\infty)$. Therefore, inserting \eq{5.21} into \eq{5.11}, we get
\begin{align}
\frac E{4\pi}&=\frac{q^2}{4\sqrt{2}a}\int^\infty_0\frac{\e^{-\frac\eta4}(1+\eta)(2+5\eta+\eta^2)}{\eta^\frac34(1+\frac\eta2)(2+\eta)^\frac32}{\rm{d}}\eta\nn\\&\approx1.50776\left(\frac{q^2}{a}\right).
\end{align}

(v) The logarithmic model \cite{Gaete2019Ospedal,Soleng1995,Feigenbaum1998,Akmansoy2018} is given by
\begin{align}
f(s)=&-\frac1\beta\ln(1-\beta s),\\
U(t)=&-\ln(1-t).\lb{5.24}
\end{align}
From \eq{5.10}, we see that the interval of the variable $\eta$ is $[0,2)$. Thus, inserting \eq{5.24} into \eq{5.11}, we arrive at
\begin{align}
\frac E{4\pi}&=\frac{q^2}{8a}\int^2_0\frac{(2+3\eta^2)(2\eta+(2-\eta)\ln(1-\frac\eta2))}{\eta^\frac32\sqrt{4-2\eta^2}}{\rm{d}}\eta\nn\\
&\approx1.38486\left(\frac{q^2}{a}\right).
\end{align}

(vi) The interpolated Maxwell--Born--Infeld theory introduced in \cite{LY} is defined by the Lagrangian action density
\begin{align}
f(s)=&\lambda s+\frac{1-\lambda}{\beta}\left(1-\sqrt{1-2\beta s}\right),\quad \lambda\in[0,1),\\ 
U(t)=&\lambda t+(1-\lambda) (1-\sqrt{1-2t}).\lb{5.27}
\end{align}
From \eq{5.10}, we see that the interval of the variable $\eta$ is $[0,1)$. Thus, inserting \eq{5.27} into \eq{5.11}, we have
\begin{align}
\frac E{4\pi}=&\frac{q^2}{4a}\int_0^1\left(\frac{\lambda }2+\frac{1-\lambda }{\sqrt{1-\eta}(1+\sqrt{1-\eta})}\right)\left(\lambda +\frac{1-\lambda }{\sqrt{1-\eta}}\right)^{-\frac52}\eta^{-\frac34}
\left(\frac{1-\lambda \eta}{(1-\eta)^{\frac32}}+\lambda +\frac{1-\lambda }{\sqrt{1-\eta}}\right)\,\dd\eta\nn\\
=&\frac{q^2}{a}H(\lambda).\lb{5.28}
\end{align}
In Table \ref{T1}, we list some results for $H(\lambda)$:
\begin{table}[htbp]
\centering
\normalsize
\renewcommand{\arraystretch}{2.2}
\setlength{\tabcolsep}{4pt}
\centering
\begin{tabular}{|c|c|c|c|c|c|c|}
\hline
{$\lambda$}&$0$&$\frac1{10}$&$\frac14$&$\frac1{2}$&$\frac34$&$\frac{9}{10}$\\
\hline
$H(\lambda)$&$1.23605$&$1.24249$&$1.25289$&$1.27275$&$1.29748$&$1.31658$\\
\hline
\end{tabular}\caption{Numerical results of the normalized energy $H(\lambda)$ of the interpolated Maxwell--Born--Infeld model.}\lb{T1}
\end{table}

These results show that in all the parameter regimes it is impossible to attain a sufficient small effective radius for an electric point charge in those models.

As a technical issue, we note that appropriate examples may be formulated which fulfill \eq{05} or \eq{5.2} but fail to observe the condition \eq{5.7} or \eq{5.9}, or both. For example, for the logistic function  
\be
f(s)=\frac4{\beta(1+\e^{-\beta s})}-\frac2\beta,
\ee
we have
\be\lb{5.31}
U(t)=\frac4{1+\e^{-t}}-2,
\ee
that fulfills \eq{5.2} of course. However, it is clear that \eq{5.31} violates both \eq{5.7} and \eq{5.9} so that the model here cannot accommodate a finite-energy electric point charge formalism.

\medskip

{\bf Robustness of the nonlinearly perturbed Maxwell theory}

We note that the generalized nonlinearly perturbed Maxwell model given by
\begin{align}
f(s)=s+\frac\gamma\beta\left(1-\sqrt{1-(2\beta s)^k}\right),\quad k=3,5,...,\quad s=-\frac14 F_{\mu\nu}F^{\mu\nu}, \label{5.36}
\end{align}
enjoys all the properties of the model \eq{1.1}. In fact,  we have
\be\lb{5.33}
U(t)=t+\gamma\left(1-\sqrt{1-(2t)^k}\right).
\ee
This function satisfies all the conditions \eq{5.2}, \eq{5.7}, and \eq{5.9}. Besides, its associated full Hamiltonian energy density reads

\begin{align}
\mathcal{H}=&\frac12(\textbf{E}^2+\textbf{B}^2)+\frac{\gamma\left(\beta[\textbf{E}^2-\textbf{B}^2]\right)^{k-1}\textbf{E}^2\left(k+(k-1)\sqrt{1-\left(\beta[\textbf{E}^2-\textbf{B}^2]\right)^k}\right)}{\sqrt{1-\left(\beta[\textbf{E}^2-\textbf{B}^2]\right)^k}\left(1+\sqrt{1-\left(\beta[\textbf{E}^2-\textbf{B}^2]\right)^k}\right)}\notag\\
&+\frac{\gamma\left(\beta[\textbf{E}^2-\textbf{B}^2]\right)^{k-1}\textbf{B}^2}{1+\sqrt{1-\left(\beta[\textbf{E}^2-\textbf{B}^2]\right)^k}}\label{x5.33}
\end{align}
which is seen not to accommodate a magnetic or, more generally, a dyonic, point charge, exactly as before for \eq{1.1}. 

By \eq{5.33}, we obtain the energy of an electric point charge as follows,
\begin{align}\lb{5.34}
\frac E{4\pi}=&\frac{q^2}aH_{k}(\gamma)\equiv \frac{q^2}{a}\int^1_0h_{k,\gamma}(\eta) {\rm{d}}\eta,
\end{align}
where the normalized Hamiltonian energy density function
\begin{align}\lb{ay6.35}
h_{k,\gamma}(\eta)=\left(1+\frac{2\gamma\eta^{k-1}(k+(k-1)\sqrt{1-\eta^k})}{\sqrt{1-\eta^k}(1+\sqrt{1-\eta^k})}\right)\frac{(\gamma\eta^{k-1}((2k-1)k-k(k-1)\eta^k)+(1-\eta^k)^\frac32)}{8\eta^\frac34(1-\eta^k)^\frac14(\gamma k\eta^{k-1}+\sqrt{1-\eta^k})^\frac52}
\end{align}
leads to the $k$-independent Maxwell theory limit
\begin{align}\lb{x5.36}
\lim_{\gamma\to0}H_{k}(\gamma)=\frac12,
\end{align}
as that when $k=3$ given by \eq{4.8}. Moreover, we have
\be\lb{5.37}
\lim_{\gamma\to\infty}H_k(\gamma)=0,
\ee
extending \eq{a4.6}. 

To evaluate the free electric charge and self energy contained in a ball around the point charge in this model, we insert \eq{5.33} into \eq{5.10} and set $r=R$ to get
\be
\left(1+\frac{\gamma k \eta^{k-1}}{\sqrt{1-\eta^k}}\right)^2\eta =\frac{a^4}{R^4},\quad \eta=\eta(\gamma),
\ee
similar as in \eq{a4.15}, with the same notation convention, resulting in the relation
\be
\gamma=\frac{\sqrt{1-\eta^k}}{k\eta^{k-\frac12}}\left(\frac{a^2}{R^2}-\eta^{\frac12}\right).
\ee
Hence \eq{ay5.20} and \eq{ay5.21} are still valid here. With \eq{ay5.21}, we see that all the scenarios (i)--(iii)  regarding the
free electric charge and the self energy contained in the ball of  radius $R$ stated in \S \ref{S4}
for the model \eq{1.1}  hold for the generalized model \eq{5.36} as well, including the local disparity phenomenon about the free electric charge and electrostatic self energy 
of an electric point charge, as spelled out by \eq{axx5.22a}, \eq{axx5.22b}, \eq{axx5.25a},  and \eq{axx5.25b}, in connection with the global energy-charge relation \eq{axx5.24}.
In fact, inserting \eq{x5.36} into \eq{5.34}, we arrive at the conclusion \eq{x4.9} again for the model \eq{5.36} in general. Furthermore, replacing \eq{4.20} for the model here, we have
\be
E(R)=\frac{4\pi q^2}a\int^1_{\eta(\gamma)} h_{k,\gamma}(\eta)\,\dd\eta.\lb{ay6.40}
\ee
Inserting \eq{ay6.35} into \eq{ay6.40} and taking $\gamma\to0$, we obtain \eq{axx5.25a} and \eq{axx5.25b} again.
\medskip

{\bf Generalized nonlinearly perturbed Maxwell theory}

Along the line of the study in our context is a generalized nonlinearly perturbed Maxwell theory that incorporates the Born--Infeld electrodynamics of the type \eq{03} and extends
\eq{5.36} and is governed by the Lagrangian action density
\begin{align}\lb{5.38}
\mathcal{L}&=-\frac14F_{\mu\nu}F^{\mu\nu}+\frac\gamma\beta\left(1-\sqrt{1-(2\beta s)^k}\right)\nn\\
&\equiv -\frac14F_{\mu\nu}F^{\mu\nu}+f(s),
\end{align}
where
\begin{align}\lb{5.39}
s=-\frac{1}{4}F_{\mu\nu}F^{\mu\nu}+\frac{\kappa^2}{32}\left(F_{\mu\nu}\tilde{F}^{\mu\nu}\right)^2
=\frac12({\bf E}^2-{\bf B}^2)+\frac{\kappa^2}2({\bf E}\cdot{\bf B})^2
\end{align}
is the sum of the usual Maxwell action density and an electromagnetic interaction term whose strength depends on the free coupling parameter $\kappa\geq0$. With \eq{5.38} and
\eq{5.39}, the associated energy-momentum tensor is given by
\begin{align}
T_{\mu\nu}\lb{5.40}
=&-F_{\mu\mu'}\eta^{\mu'\nu'}F_{\nu\nu'}+\frac14g_{\mu\nu}\left(F_{\mu'\nu'}F^{\mu'\nu'}\right)\nn\\
&-f'(s)\left(F_{\mu\mu'}\eta^{\mu'\nu'}F_{\nu\nu'}-\frac{\kappa^2}4(F_{\mu'\nu'}\tilde{F}^{\mu'\nu'})F_{\mu\mu''}\eta^{\mu''\nu''}\tilde{F}_{\nu\nu''}\right)-\eta_{\mu\nu}f(s).
\end{align}
Thus, by virtue of \eq{5.38}--\eq{5.40}, the induced Hamiltonian energy density ${\cal H}=T_{00}$ is given by
\begin{align}
\mathcal{H}=\frac12(\textbf{E}^2+\textbf{B}^2)+\frac{\gamma k\left(2\beta s\right)^{k-1}\textbf{B}^2}{1+\sqrt{1-\left(2\beta s)\right)^k}}
+\frac{\gamma\left(2\beta s\right)^{k-1}(\textbf{E}^2+\kappa^2(\textbf{E}\cdot\textbf{B})^2)\left(k+(k-1)\sqrt{1-\left(2\beta s\right)^k}\right)}{\sqrt{1-\left(2\beta s\right)^k}\left(1+\sqrt{1-\left(2\beta s\right)^k}\right)},\label{5.41}
\end{align}
which is apparently positive definite when $k$ is an odd integer,  $k=3,5,7...$,  and recovers \eq{x5.33} when setting $\kappa=0$.

In view of \eq{5.41},  we conclude that a magnetic point charge of the form \eq{a3.20} leads to energy divergence. Thus, the generalized nonlinearly perturbed Maxwell theory
consisting of \eq{5.38} and \eq{5.39} does not accommodate a monopole or dyon due to the finite-energy condition. As a consequence, in a point charge situation, only an electric 
point charge is energetically accepted, such that the theory returns to that governed by the simplified model \eq{5.36} where the quantity $s$ is defined by the Maxwell action
density given in \eq{1.1}, or \eq{5.39} with setting $\kappa=0$.

\section{Polynomial model and Maxwell theory limit}\label{S6}

The study of the previous two sections shows that the self energy of an electric point charge in the Maxwell theory limit, $\gamma\to0$, remains finite, despite that
energy divergence takes place at $\gamma=0$ in the context of the Maxwell theory.  Specifically, for an electric point charge, the theory \eq{1.1}, or more generally, \eq{5.36}
or \eq{5.38}--\eq{5.39}, does not return to the Maxwell theory (at $\gamma=0$)  in the Maxwell theory limit (as $\gamma\to0$). This phenomenon seems rather surprising and
interesting and raises the question whether there is a nonlinearly perturbed Maxwell theory of the type
\be\lb{x7.1}
{\cal L}= -\frac14F_{\mu\nu}F^{\mu\nu}+\gamma p(s),\quad \gamma\geq 0,
\ee
where $s$ is defined by \eq{5.39} and $p(s)$ is nonlinear function of $s$ satisfying $p(0)=0$ and $p'(0)=0$, which enjoys the desired properties

\begin{enumerate}

\item[(i)] A finite-energy electric point charge is accommodated but not a monopole or dyon.

\item[(ii)] The parameter $\gamma$ can be adjusted such that the effective radius of an electric point charge may be made as small as possible.

\item[(iii)] The free charge contained in the ball of a sufficiently large radius around an electric point charge approaches the prescribed point charge as $\gamma\to0$.

\item[(iv)] The self energy of an electric point charge blows up as $\gamma\to0$ to preserve the Maxwell theory limit.

\end{enumerate}

We have considered this problem and found that any polynomial of the form
\be\lb{x6.2}
p(s)=\sum_{k=2}^n a_k s^k,\quad a_k\geq0,\quad k=2,\dots,n,\quad a_2+\cdots+a_n>0,
\ee
in \eq{x7.1}, subject to an appropriate reparametrization,  achieves the purpose. Specifically, for this purpose, we introduce the parameter $\beta$ in \eq{x6.2} 
such that $a_k=b_k\beta^{k-1}$ ($k=2,\dots,n$) to recast \eq{x6.2}  into
\be\lb{6.1}
p(s)=\sum_{k=2}^n b_k\beta^{k-1}s^k,
\ee
so that \eq{x7.1} becomes \eq{5.2} with $s$ given in \eq{1.1} and the function $U$ formulated in \eq{5.2} reads
\be\lb{6.2}
U(t)=t+\gamma \sum_{k=2}^n b_k t^k.
\ee
It is obvious that \eq{5.7} is valid. Besides, \eq{5.9} also holds since
\be
2U'(t)t-U(t)=t+\gamma\sum_{k=2}^n (2k-1) b_k t^k\geq0.
\ee

On the other hand, as a consequence of the expression \eq{5.40} where $f(s)=\gamma p(s)$ with $p(s)$ given in \eq{6.1} and $s$ defined in general by \eq{5.39}, we see that, in the point-charge situation, monopoles and dyons are again
energetically excluded, and the system only allows an electric point charge. In such a situation, we return to the theory \eq{6.2} and the associated
Hamiltonian energy density spelled out by \eq{5.8}  is
\begin{align}
    \mathcal{H}=\frac{1}{2}\textbf{E}^2+\gamma\sum^n_{k=2}\frac{(2k-1)a_k}{2^k}(\textbf{E}^2)^k,\label{x7.6}
\end{align}
and  the constitutive equation \eq{5.10} renders us the expression
\begin{align}
\frac{a^4}{r^4}=\left(1+\gamma\sum^n_{k=2}k b_k \left(\frac\eta2\right)^{k-1}\right)^2\eta,\label{x7.7}
\end{align}
giving rise to the asymptotic relations between the variables $\eta$ and $r$:
\be
r\to0,\quad \eta\to\infty;\quad \eta\sim r^{-\frac4{2n-1}},\quad r\ll1;\quad r\to\infty,\quad \eta\to0;\quad \eta\sim r^{-4},\quad r\gg1.
\ee
Therefore, substituting \eq{6.2} into \eq{5.11} and setting $\eta=2t$ for convenience of calculation, we have 
\begin{align}
\frac E{4\pi}=&\frac{q^2}{a}H_n(\gamma)\equiv\frac{q^2}a\int_0^\infty h_{n,\gamma}(t)\,\dd t\nn\\
=&\frac{q^2}a\int_0^\infty
\frac{(1+\gamma \sum_{k=2}^n k(2k-1)b_k t^{k -1}
)
(1+\gamma \sum_{k=2}^n (2k-1)b_k t^{k -1})}
{2^{\frac{11}4} t^{\frac{3}{4}} \left(1+ \gamma \sum_{k=2}^n k b_k t^{k-1} \right)^{\frac{5}{2}}}\,\dd t\label{6.5}
\end{align}
which gives us the limits
\begin{align}
\lim_{\gamma\to0} H_n(\gamma)=&\infty,\lb{6.6}\\
\lim_{\gamma\to\infty}H_n(\gamma)=&0.\lb{6.7}
\end{align}
In fact, \eq{6.7} follows from \eq{6.5} obviously, which gives rise to the result that the effective radius of an electric point charge can be made arbitrarily small, as before, and \eq{6.6}
is seen from using Fatou's lemma in \eq{6.5}: 
\be
\liminf_{\gamma \to0}H_n(\gamma )\geq \int_0^\infty \liminf_{\gamma\to0}h_{n,\gamma}(t)\,\dd t=\int^\infty_0\frac{\rm{d}\eta}{2^{\frac{11}4}\eta^\frac34}=\infty. \lb{6.8}
\ee

The property \eq{6.8} is indeed compatible with what happens at the Maxwell theory limit, $\gamma=0$. Technically, this divergence occurs at $t=\infty$ corresponding to $r=0$, hence realizing an ultraviolet divergence as in the Maxwell theory. Thus, energetically, the limit $\gamma\to0$
recovers the Maxwell theory, unlike what happens in the model \eq{5.36}, in sharp contrast.

\begin{figure}[htbp]
\centering
    \includegraphics[width=0.6\linewidth]{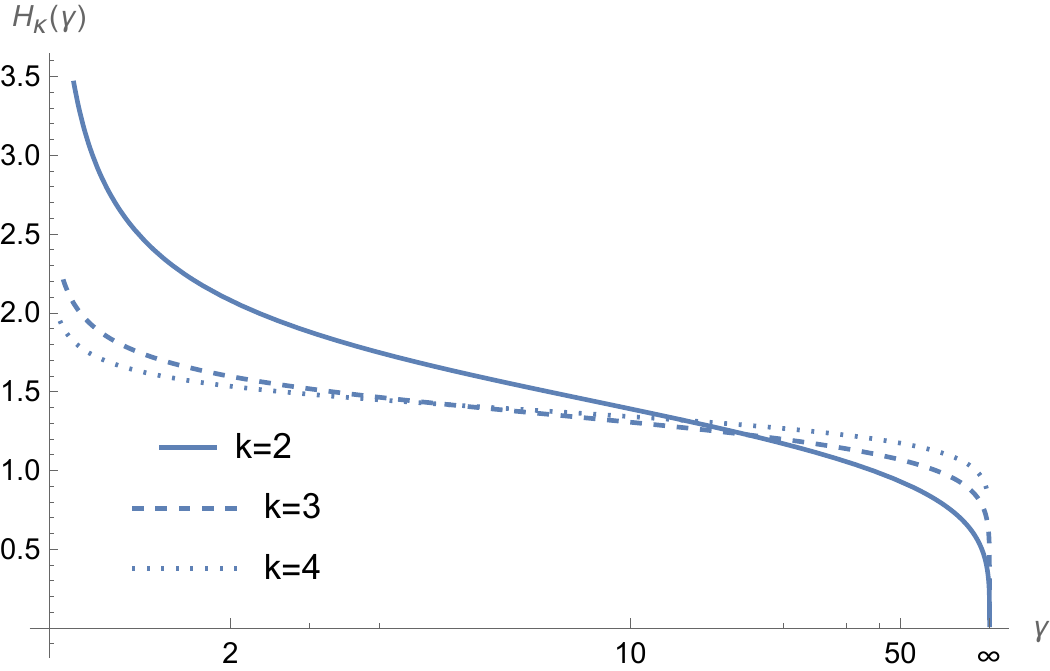}
    \caption{\footnotesize{Plots of the normalized energy $H_k(\gamma)$  associated with the monomial model \eq{7.13} for $k=2,3,4$. The divergence of $H_k(\gamma)$ as
$\gamma\to0$ indicates that the model energetically recovers the Maxwell theory in describing an electric point charge and the limit $H_k(\gamma)\to0$ as $\gamma\to\infty$ implicates that
the effective radius of an electric point charge can be made arbitrarily small when choosing $\gamma$ sufficiently large. }}
    \label{F4}
\end{figure}

In Figure \ref{F4}, we present the numerical results of the normalized energy $H_k(\gamma)$ for $k=2,3,4$, for the monomial model
\be\lb{7.13}
p(s)= \beta^{k-1} s^k,
\ee
confirming the asymptotic properties
as stated in \eq{6.6} and \eq{6.7}.

\medskip

{\bf Free electric charge distribution}

Now return to the polynomial model \eq{6.2}.

 In view of \eq{xx5.15}, the normalized free electric charge contained in the ball of  radius $R$ around the point charge is given by
\be\lb{7.14}
q_{\rm{free}}(R)=\frac{Q_{\rm{free}}(R)}{4\pi}=q\left(\frac{R}{a}\right)^2 \eta^{\frac12},
\ee
where $\eta=\eta(\gamma)$ is determined by \eq{x7.7} with setting $r=R$, which may be resolved to yield
\be
\gamma=\frac{\frac{a^2}{R^2}-{\eta}^{\frac12}}{\sum_{k=2}^n \frac{k b_k}{2^{k-1}}\eta^{k-\frac12}},
\ee
which decreases with respect to $\eta\in\left(0,\frac{a^4}{R^4}\right]$ and enjoys the asymptotic properties
\be\lb{7.16}
\eta(\gamma)\to\frac{a^4}{R^4},\quad \gamma\to0;\quad \eta(\gamma)\to0,\quad \gamma\to\infty.
\ee
Using \eq{7.16} in \eq{7.14}, we arrive at the limits 
\bea
&& \lim_{\gamma\to0} q_{\rm{free}}(R)= q,\lb{ax7.17}\\
&&\lim_{\gamma\to\infty} q_{\rm{free}}(R)=0,\lb{ax7.18}
\eea
for any $R>0$. It is interesting to compare \eq{ax7.17} with \eq{axx5.22a} and \eq{axx5.22b} and observe that \eq{ax7.17} is true for arbitrary radius $R$, which indicates that we indeed
return to the Maxwell theory
in the zero nonlinearity limit electrically for a point charge. Meanwhile, the same vanishing result, \eq{ax7.18} continues to be valid in the strong nonlinearity limit as that for the 
non-polynomial model \eq{5.36} as stated in (ii) in \S \ref{S4}.

\medskip

{\bf Energy distribution}

It will also be interesting to consider the electrostatic self energy contained within a ball of any radius around a point charge and compare it with that in the full space in the limiting situations
as $\gamma\to0$ and $\gamma\to\infty$, respectively. In fact, with the notation \eq{6.5}, we have
\be
\frac{E(R)}{4\pi}=\frac{q^2}a \int_{\eta(\gamma)}^\infty \frac{(1+\gamma \sum_{k=2}^n k(2k-1)b_k t^{k -1}
)
(1+\gamma \sum_{k=2}^n (2k-1)b_k t^{k -1})}
{2^{\frac{11}4} t^{\frac{3}{4}} \left(1+ \gamma \sum_{k=2}^n k b_k t^{k-1} \right)^{\frac{5}{2}}}\,\dd t.\label{7.17}
\ee
Thus, in view of \eq{7.16} and Fatou's lemma, we have
\be\lb{7.18}
\lim_{\gamma\to0}E(R)=\infty.
\ee
Hence the Maxwell type ultraviolet divergence is a local phenomenon as expected.

On the other hand, since $E(R)\to 0$ as $\gamma\to\infty$, we see that the limit of the ratio
\be\lb{7.19}
\rho(R)\equiv\frac{E(R)}E,
\ee
as $\gamma\to\infty$, is of the zero-over-zero type in both models, \eq{5.36} and \eq{x7.1}, which is difficult to evaluate in general. In Table \ref{tab1}, we present 
a collection of numerical results of the ratio \eq{7.19} for the model \eq{7.13} with $k=3$  and $R=a$ for some values of $\gamma$. These results show that the ratio \eq{7.19} decreases
as $\gamma$ increases and that it converges to a positive limit as $\gamma\to\infty$.
\begin{table}[htbp]
    \centering
    \large
    \begin{tabular}{c|c|c|c|c}
        $\gamma$&10&100&1000&$10^6$  \\
        \hline
         $\eta(\gamma)$&0.14587&0.039291&0.009303&0.0001\\
         \hline
         $\rho(a)$&0.62544&0.542927&0.464934&0.2777
    \end{tabular}
    \caption{A table of a few numerical sample results of  the energy ratio $\rho(a)=\frac{E(a)}{E}$ of the energy of an electric point charge contained in
the ball of the effective radius of the point charge against that distributed in the full space in the strong nonlinearity limit.}
    \label{tab1}
\end{table}
 
A similar study has been carried out for the model \eq{1.1} which shows that the ratio \eq{7.19} approaches zero as $\gamma\to\infty$. This result is of particular interest since
it demonstrates that, as $\gamma\to\infty$, the electric point charge is locally and relatively undetectable {\em both} electrically and energetically.

\section{Conclusions and comments}\label{S7}

In this work we have pursued a classical question at the heart of nonlinear electrodynamics: how to reconcile the pointlike idealization of the electron with a finite self–energy requirement and a physically meaningful effective radius. Following the tradition of Born and Infeld \cite{B1,B2,BI1,BI2}, we proposed a new class of models that are nonlinear perturbations of Maxwell’s theory rather than fully nonlinear deformations of the theory, as that of Born and Infeld.

Building upon the models \eq{5.36},  \eq{5.38}, and \eq{x7.1}, with non-polynomial and polynomial perturbations, respectively,
our central finding is that, through a judicious choice of a coupling parameter, the effective radius of an electric point charge can be made arbitrarily small, for example, well below the unattainable $10^{-20}\,\mathrm{cm}$ scale in the Born--Infeld construction for the electron. At the same time, the self energy remains finite and can be apportioned in a controlled way between the effective radius ball and the surrounding space. These observations crystallize into the following principles which we state as theorems along with some comments.

\begin{theorem} {\rm (Effective Radius Reduction)} In the nonlinearly perturbed Maxwell models of the form introduced here, the effective radius $a$ of an electric point charge of mass $m$ and charge $q$ satisfies
\be
a = \frac{q^2}{m}\, H(\gamma),
\ee
where $H(\gamma)$ is a normalized dimensionless energy functional depending on a dimensionless coupling parameter $\gamma$, given in the models. As $\gamma \to \infty$, $H(\gamma) \to 0$, allowing the effective radius to shrink arbitrarily, while preserving a finite self energy.
\end{theorem}

In the work of Born and Infeld \cite{BI1,BI2}, the effective radius $a$ as stated in \eq{04} or
\eq{axx5.3} of an electric point charge arises as a foundational length scale conceptually analogous to the Planck length and Fermi temperature but has not been given an interpretation
as how {\em effective} this length scale is in capturing the electric ingredient of an electric point charge. Our study enables us to obtain a signature interpretation of such a length scale
which we now state as follows.

\begin{theorem} {\rm (Signature Interpretation of Effective Radius)}
For the non-polynomial perturbed models \eq{5.36} and  \eq{5.38}, the normalized free electric charge generated by the electric field of an electric point charge of the charge $q>0$ enjoys the universal property
\be\lb{8.2}
 \lim_{\gamma\to0} q_{\rm{free}}(R)=q\left(\frac Ra\right)^2,\quad R\leq a;\quad \lim_{\gamma\to0} q_{\mathrm{free}}(R)=q,\quad R\geq a,
\ee
 in the Maxwell theory limit. That is, in the Maxwell theory limit, the free electric charge contained in the ball of radius $R$ around the point charge attains its maximum, which is the total prescribed point charge,
at the minimum radius $a$,  which is the effective radius of the point charge. In other words, 
the effective radius indeed truly effectively captures all possible charge within its reach. Besides its electric characterization, the effective radius $a$ also plays the role of a critical scale below which the self energy of a point charge vanishes in the Maxwell theory limit:
\be
\lim_{\gamma\to 0} E(R)=0,\quad R\leq a;\quad
 \lim_{\gamma\to 0} E(R)=2\pi q^2\left(\frac1a-\frac1R\right),\quad R\geq a,
\ee
which indicates that an electric point charge is locally energetically unobservable in Maxwell's limit.
\end{theorem}

We emphasize that such a signature {\em critical} property of the effective radius  is not available for the polynomially perturbed model \eq{x7.1} because \eq{ax7.17} is true for any $R>0$.

In sharp contrast to \eq{8.2} and \eq{ax7.17}, we have shown that \eq{ax7.18} is true in all situations. This result could explain why the electron is locally invisible as
a point charge stated as follows.

\begin{theorem}\lb{th8.3} {\rm (Electric and Energetic Invisibility of  Point Charge of  Small Effective Radius)}
 For any $R>0$ and in all model situations,  the free electric charge and the associated electrostatic energy contained in the ball of  any radius $R>0$ centered about the point charge enjoy the properties
\be
\lim_{\gamma\to\infty} q_{\mathrm{free}}(R)=0, \quad \lim_{\gamma\to\infty} E(R)=0.
\ee
Since the limit $\gamma\to\infty$ is equivalent to the limit $a\to0$ in all models, we conclude that an electric point charge becomes locally electrically and energetically undetectable in the strong nonlinear strength limit
or the zero effective radius limit.
\end{theorem}

As a by-product of our study,  we emphasize that an equally striking property of our models is the exclusion of monopoles and dyons: The leading Maxwell theory term in the action density
in the models formulated here enforces an energetic barrier that naturally rules out finite-energy magnetic or dyonic point charges. This exclusion principle parallels our earlier findings \cite{LY} in interpolated Maxwell--Born--Infeld theories and appears to be a robust structural feature of the formalism.  We summarize this principle as follows.

\begin{theorem} {\rm (Exclusion of Monopoles and Dyons)} In all the models considered, the linear Maxwell action density term imposes a natural mechanism which rules out 
finite-energy monopoles and dyons but the nonlinear perturbations serve to accommodate finite-energy electric point charges.
\end{theorem}

We have seen that, in comparison with the non-polynomial models, the polynomial models display a complementary behavior: While still enabling arbitrarily small effective radii, they  recover the Maxwell theory in the $\gamma \to 0$ limit by allowing the self energy to diverge in the expected ultraviolet manner. Thus, the polynomial family restores the classical Maxwellian asymptotics while retaining nonlinear regularization at finite coupling.

From a broader perspective, the constructions here not only shed light on the century-old puzzle of modeling the electron as a pointlike classical particle but also enrich the landscape of admissible nonlinear theories of electromagnetism. The concurrence of finite self-energy electric point charges, arbitrarily small effective radii and their 
electric and energetic characterizations, interpretation of undetectability of pointlike charges, and exclusion of monopoles suggests that nonlinear perturbations of Maxwell’s equations may provide a fertile mathematical framework for reconciling point-particle idealizations with physical consistency.

We emphasize, however, that this classical-field explanation complements rather than replaces quantum arguments: The present mechanism demonstrates how nonlinear constitutive relations can render a pointlike source effectively invisible at small scales by moving the measurable charge distribution outward. Whether and how such classical redistribution interfaces with quantum field theoretic effects (renormalization, radiative corrections, etc.) remains an interesting direction for further study.

Future work may explore the dynamical aspects of these models, their coupling with gravity in black hole contexts, or their role in effective field theories inspired by string theory and cosmology. In this sense, the present investigation may be viewed as a step in the ongoing dialogue between classical field theory, particle modeling, and high-energy physics, continuing the line of thought first set in motion by Born and Infeld nearly a century ago.

From a technical viewpoint and as a by-product, our study also sharpens and refines the definition of generalized nonlinear electrodynamics by subjecting it to a collection of minimally imposed  conditions, \eq{5.2}, \eq{5.7}, and
\eq{5.9}, for which \eq{5.2} is set such that the Maxwell theory is recovered in the weak field limit, \eq{5.7} ensures the invertibility of the electric constitutive equation relating the electric displacement field to the electric field, and \eq{5.9} is for the energy density of the induced electric field to stays non-negative. These conditions should be observed as initial sufficient conditions for
the foundation for a generalized model of electromagnetism within the spirit of the Born--Infeld formalism.

\medskip

{\bf Physical interpretation and mechanism of experimental invisibility}

A central result of this study is the realization of the pointlike electron through the limit of vanishing effective radius ($a \to 0$, or equivalently $\gamma \to \infty$). While classical Maxwellian theory yields an effective radius that contradicts high-energy scattering data, our model demonstrates that $a$ should not be interpreted as a fixed, fundamental constant of a rigid charged body. Instead, $a$ functions as a phenomenological scale or a regularization indicator that dictates the transition between the classical Maxwellian regime and the limit of experimental invisibility.

Historically, the ``size" of an electron has been a source of tension between theory and measurement. As shown, Theorem \ref{th8.3} provides a rigorous bridge across this gap. In the strong-coupling limit ($\gamma \to \infty$), the charge and energy contained within any finite neighborhood of the point source vanish. This does not imply a loss of physical identity; rather, it provides a field-theoretic mechanism for why the electron is perceived as a structureless point in every experiment to date. In this regime, the effective radius $a$ serves to {\em fine-tune} the locality of the field: As $a$ shrinks below the resolution of current probes (approximately $10^{-20}$ cm), the particle becomes effectively invisible to local detection while maintaining its global identity through the conservation of total charge $q$ and rest mass $m$ in the full space.

Consequently, the {\em discrepancy} noted by classical models is resolved here not by proposing a specific measurable ball size, but by proving that a sufficiently small $a$ renders the electron’s internal structure {\em unobservable}. The parameter $a$ is thus well-motivated as a classical distribution radius when $\gamma \to 0$, yet it remains physically consistent as an invisibility threshold in the high-energy limit, aligning the theory with the experimental null results that define modern particle physics.

In high-energy physics, the idea that a particle’s internal structure can remain ``invisible" to experiments is a well-established concept. For instance, in quantum chromodynamics, quarks behave as structureless point particles when probed at extremely high energies, a phenomenon known as asymptotic freedom, even though they are part of a complex, strongly interacting system. Similarly, in the standard theory of electrons (quantum electrodynamics), a bare electron is thought to be surrounded by a cloud of virtual fluctuations that effectively screens its properties. These parallels suggest that pointlike behavior is often an emergent feature defined by the limits of our measurement technology. In this sense, the work here  provides a classical bridge to those quantum ideas, demonstrating that a particle can maintain a full-space field presence (total mass and charge) while remaining locally undetectable, effectively hiding its internal complexity from the most advanced probes.

\medskip

{\bf Data availability statement}: The data that supports the findings of this study are
available within the article.

{\bf Conflict of interest statement}:
The authors declare that they have no known competing financial interests or personal relationships that could have appeared to influence the work reported in this article.

\end{document}